\documentclass[useAMS,usenatbib,a4]{mn2e}  

\usepackage[dvips]{graphicx}
\usepackage{natbib}
\usepackage{multirow}

\title[Stellar evolution uncertainties and UV lines in LBGs]{The effect of
  stellar evolution uncertainties on the rest-frame ultraviolet stellar lines of CIV and HeII
  in high-redshift Lyman-break galaxies} 

\author[Eldridge \& Stanway]{John J. Eldridge$^{1}$ \thanks{E-mail: jje@ast.cam.ac.
uk} \& Elizabeth R. Stanway$^{2,3}$  \\
$^{1}$Institute of Astronomy, The Observatories, University of Cambridge, Madingley Road, Cambridge, CB3 0HA, UK\\
$^{2}$H H Wills Physics Laboratory, Tyndall Avenue, Bristol, BS8 1TL, UK\\
$^{3}$Department of Physics, University of Warwick, Gibbet Hill Road, Coventry, CV4 7AL, UK\\}

\pagerange{\pageref{firstpage}--\pageref{lastpage}} \pubyear{2011}
\begin{document}
\maketitle
\label{firstpage}

\begin{abstract}
Young, massive stars dominate the rest-frame ultraviolet spectra of
star-forming galaxies. At high redshifts ($z>2$), these rest-UV
features are shifted into the observed-frame optical and a combination
of gravitational lensing, deep spectroscopy and spectral stacking
analysis allows the stellar population characteristics of these
sources to be investigated. We use our stellar population synthesis
code {\sc BPASS} to fit two strong rest-UV spectral features in
published Lyman-break galaxy spectra, taking into account the effects
of binary evolution on the stellar spectrum. In particular, we
consider the effects of quasi-homogeneous evolution (arising from the
rotational mixing of rapidly-rotating stars), metallicity and the
relative abundance of carbon and oxygen on the observed strengths of
He\,{\sc II} $\lambda$\,1640\AA\ and C\,{\sc IV}
$\lambda$\,1548,1551\AA\ spectral lines. We find that Lyman-break
galaxy spectra at $z\sim2-3$ are best fit with moderately sub-solar
metallicities, and with a depleted carbon-to-oxygen ratio. We also
find that the spectra of the lowest metallicity sources are best fit
with model spectra in which the He{\sc II} emission line is boosted by
the inclusion of the effect of massive stars being spun-up during
binary mass-transfer so these rapidly-rotating stars experiencing
quasi-homogeneous evolution.
\end{abstract}

\begin{keywords}
  galaxies: starburst -- galaxies: stellar
  content -- binaries: general -- stars: evolution -- stars:
  Wolf-Rayet
\end{keywords}

\section{Introduction}
\label{sec:introduction}

Massive stars dominate the light and kinematics of young and
star-forming galaxies. They spend most of their lifetimes with
effective temperatures greater than $10^4$\,K, and hence most of their
light is emitted in the ultraviolet (UV) region of the electromagnetic
spectrum. It is difficult to observe emission in the far-UV
($\sim$1000-2000\,\AA) from sources in the local Universe, due to
absorption in the Earth's atmosphere and the limitations of
space-based instruments sensitive to this wavelength range. At
redshifts above two this emission moves into the range of optical
spectroscopy. The study of rest-frame UV spectral lines therefore
become relatively straightforward and provides a direct insight into
the massive stellar population of star-forming and
ultraviolet-luminous galaxies.

However a problem arises from the extreme distance and consequent
apparent faintness of galaxies at such redshifts. The low
signal-to-noise obtained in observations of typical individual
galaxies makes their line strength and stellar population properties
difficult to determine with any degree of precision. One approach to
circumventing this problem and characterising a galaxy population is
to combine a number of spectra to obtain a single, mean composite
spectrum.  Studies of such stacked spectra, notably the composite
constructed by \citet{shapley} of $z\sim3$ rest-UV selected
`Lyman-break galaxies' (LBGs), allow the determination of absorption
and emission line intensities that are not detectable in individual
sources, while being limited to yielding information on a `typical'
source rather than any individual galaxy.

However, the analysis of \citet{shapley} revealed problems with our
understanding of the stellar populations in such systems. The standard
stellar population models produced by \textit{Starburst99}
\citep{1999ApJS..123....3L} were not able to simultaneously reproduce
the observed C\,{\sc IV} $\lambda$\,1548,1551\AA\ stellar P-Cygni
absorption feature and He\,{\sc II} $\lambda$\,1640\AA\ stellar
emission line. The shape and strength of these broad lines are
determined primarily by the stellar content of a galaxy, in
combination with a relatively narrow nebular component. Massive
main-sequence OB stars dominate the C\,{\sc IV} line (with a small
contribution from highly stripped WC Wolf-Rayet stars), while the
He\,{\sc II} line emission primarily arises from evolved, hydrogen
poor, Wolf-Rayet (WR) stars (with a small contribution from rare but
luminous Of stars). Recently \citet{brinchmann1} revised predictions
for the He\,{\sc II} line strengths expected from a stellar population
by including updated Wolf-Rayet star line equivalent widths (EWs) in
the \textit{Starburst99} population synthesis code. From this the
authors were able to predict a He\,{\sc II} EW that was in agreement
with that observed in the \citet{shapley} spectrum, suggesting that
careful consideration of these massive stars and their evolutionary
pathways is vital in the interpretation of these distant sources.

While the use of composite spectra is necessary to study the majority
of distant sources, there are now a small but growing number of
individual Lyman-break galaxies for which high signal-to-noise rest-UV
spectra have been obtained. The majority of these observations are made possible by
high magnification due to strong lensing by intervening material along
the line of sight.  They have highlighted a diversity in the
population that is obscured by use of a composite, with some galaxies
having extremely strong He\,{\sc II} emission and others vanishingly
weak, undetectable lines. This variety leads to a dilemma in its
interpretation. While strong C\,{\sc IV} lines indicate the presence
of massive O stars, a weak He\,{\sc II} line would suggest a
surprising absence of WR stars in the same galaxies. This dichotomy
suggests that a reexamination of the spectral synthesis predictions of
these C\,{\sc IV} and He\,{\sc II} features is necessary.

Much work has been done over the past few years improving stellar
population and spectral synthesis and quantifying the uncertainties of
these models
\citep[e.g.][]{1999ApJS..123....3L,2003MNRAS.344.1000B,2004A&A...425..881L,2005MNRAS.362..799M,2007ASPC..374..303B,uncertain,uncertain1,uncertain2}.
Our approach has been to consider the uncertainties inherent in the
use of only single-star evolution models as the building blocks of
synthesis codes. We have expanded our input stellar evolution models
to take account of binary stars and the many new evolutionary paths
ways this entails. In \citet{ES09} we showed that when recent single
star stellar models are used to predict the He\,{\sc II} line
strength, the measured line equivalent width (EW) is very small, and
that binary evolution models are a better match for the observed
He\,{\sc II} strength (and for other spectral WR population
indicators) in star-forming galaxies.

Here we build on that work, and consider two uncertainties in the
stellar evolution models in an effort to interpret both the
\citet{shapley} composite spectrum and those of individual lensed LBGs
at $z=2-3$. First, we investigate the effect of variation in the
surface carbon abundance of OB stars on the derived stellar
spectra. \citet{erb} show that the amount of carbon in high redshift
galaxies, as measured by nebular emission lines, decreases more
rapidly than their [O/H]-derived metallicity at low
abundances. Therefore we investigate the effect of decreasing the
relative abundance of carbon on the C\,{\sc IV} absorption line, in
order to examine and model this trend.

Second, we further investigate the importance of duplicity in a
synthesised stellar population when determining the He\,{\sc II} line
equivalent width, assuming both instantaneous burst and constant
star-formation histories. We now include a new evolutionary process
that may only be possible at low metallicities. Quasi-homogeneous
evolution (QHE) occurs when a massive star rotates sufficiently
rapidly that the star becomes fully mixed during its main-sequence
lifetime \citep{maeder87,yoon1,2007A&A...464L..11M}. While it is
difficult for a single star to be born rapidly rotating, efficient
mass-transfer in massive binaries can spin up secondary stars to high
rotation rates with ease \citep{cantiello}. At Solar metallicities
such stars are likely to spin down quickly due to strong stellar
winds. However with the weakening of stellar winds at lower
metallicities, efficient rotational mixing can occur. In this paper we
model the effect of this unusual form of evolution on stars that
accrete material during a mass-transfer event, and investigate the
dramatic effect such stars would have on the integrated spectrum of a
stellar population.

This paper is organised as follows. In Section
\ref{sec:synthetic_spectra} we discuss the modifications made to our
Binary Population and Spectral Synthesis (BPASS) code in order to
investigate the phenomena outlined above. In Section
\ref{sec:pred-equiv-widths} we describe the effects of these phenomena
on the strength of the diagnostic He\,{\sc II} and C\,{\sc IV}
spectral features. In Section \ref{sec:comp_spect} we go on to compare
our predicted spectra to a composite spectrum constructed from
$\sim$900 Lyman-break galaxies by \citet{shapley} and to various
individual examples of $z\sim2-3$ Lyman-break
galaxies. Finally, in Section \ref{sec:conclusions} we briefly discuss
our results and outline our conclusions.

\section{Synthetic spectra of stellar populations}
\label{sec:synthetic_spectra}

\subsection{Binary
Population and Spectral Synthesis (BPASS)}
\label{sec:bpass}

The synthetic spectra used in this paper are created using the Binary
Population and Spectral Synthesis (BPASS)
code\footnote{\texttt{http://www.bpass.org.uk}}. It is
described in detail in \citet{EIT}, \citet{ES09} and
\citet{2011arXiv1103.1877E}.

We use stellar models from the Cambridge STARS code \citep[][ and
  references therein]{egg,EIT}, specifically those calculated in
\citet{EIT}. Their key feature is an extensive set of detailed binary
star models (in addition to the detailed single star models) which are
key to producing a realistic synthetic stellar population. We consider
stellar models at five different metallicities: $Z=0.001$, 0.004,
0.008, 0.020 and 0.040 (where a metallicity of $Z=0.020$ is
conventionally considered Solar), with hydrogen mass fraction,
$X=0.75-2.5Z$, helium mass fraction, $Y=0.25+1.5Z$ and a default metal
distribution given by scaled-Solar abundances.

Given that stellar evolution is non-linear and binary evolution is
even less predictable, we do not interpolate between models with
different masses and initial binary parameters, but rather weight each
stellar model by an initial mass function (IMF) and distribution of
binary properties. Binary population fractions, pathways and mass loss
rates are calculated as described in \citet{EIT} and \citet{ES09}. The
details of our binary interaction algorithm are relatively simple
compared to the scheme outlined in, for example, \citet{HPT02} and
interactions depend primarily on the initial separation and stellar
mass ratio.  Our aim was to investigate the effect of enhanced mass
loss due to binary interactions on stellar lifetimes and populations;
therefore we concentrated on these aspects rather than incorporating
additional physical processes, each of which would add more free
parameters to our models and potentially associated uncertainties on
those parameters or the mechanisms concerned.

\subsection{Quasi-homogeneous evolution}
\label{sec:QHE}

We include one new evolutionary path, not in the standard picture of
binary evolution for our secondary stars, first introduced into our
population synthesis and described in detail in
\citet{2011arXiv1103.1877E}. This path only occurs if the more massive
primary star in a binary overfills is Roche Lobe and transfers mass
to the secondary star. If this happens and the secondary has a
metallicity mass fraction less than or equal to $0.004$, a mass after
accretion of more than 10\,M$_{\odot}$ and has accreted more than 5
percent of its initial mass, we assume it evolves fully mixed during
its main-sequence lifetime \citep{petrovic05,cantiello}. This is
referred to as quasi-homogeneous evolution and is the result of rapid
rotation due to the accretion of material from the primary star as
described by \citet{yoon1} \citep[see
  also][]{maeder87,2007A&A...464L..11M,yoon2,cantiello}.

In this case we use simple models in which we assume that stars
meeting the criteria for QHE are fully mixed during their hydrogen
burning evolution. They burn all their hydrogen to helium and increase
in surface temperature as the burning progressing, evolving away from
the zero-age main-sequence \textit{in the wrong direction}. Once
hydrogen burning ends this mixing also comes to an end. This evolution
prescription is applied whether or not the binary remains bound after
the primary supernova. The stars never grow to large radii, because
the homogeneous evolution means the increasing molecular weight of the
stellar material makes the stars shrink throughout their main-sequence
evolution. Therefore QHE stars in a binary never fill their Roche lobe
and interact with the other binary component. Other than this
inclusion of mixing, the stars are treated identically to the non-QHE
models with the same mass-loss rates applied.

\subsection{Accounting for relative carbon abundance}\label{sec:vary-carbon}

In our standard models, and those of other population synthesis codes,
Solar-scaled relative abundances are normally employed for a range metals.
Variation in the relative abundance of different species is also
possible - for example, an enhanced abundance of elements generated by
the alpha process is characteristic of bursty, massive star formation.
Recent studies modelling the interstellar medium in some high redshift
galaxies have suggested that the important ratio of carbon to oxygen
abundance decreases along with the total metallicity of the galaxies
\citep[see][for example]{erb}.  These abundance variations must be
considered in order to develop an accurate model for stellar
population synthesis. However, adapting the synthetic stellar
populations is far from straightforward. For a full analysis, the
stellar models and atmosphere models should all be recreated with
freely-varying relative abundances. This would be a challenging and
time consuming task, as well as providing many additional free
parameters which are likely to be unconstrained by the data. In this
paper we outline a simple prescription for estimating the effect of
varying carbon/oxygen ratios on the synthetic spectrum predicted by a
stellar population in the UV.

In addition to variations in the initial carbon/oxygen ratios, the
relative abundance of different elements in a stellar interior changes
over time due to nuclear reactions. While these reactions occur at the
core of the star, various processes can make nuclear-processed
material observable at the surface of a star. For stars above a few
Solar masses main-sequence hydrogen burning is catalysed by CNO
elements. During this catalysis the rate determining step is a decay
of $^{14}$N and so most of the initial CNO abundance ends up in this
element. This abundance change is concentrated at the core of the
star, but mass-loss of the outer layers and additional mixing (beyond
that predicted by mixing-length theory) can draw the the processed
core material to the surface thus changing the observable
abundance. Typically this is observed as nitrogen enrichment and used
to infer the importance of rotational mixing in stellar interiors
\citep{flames,2009A&A...496..841H}.

To take account of a possible decrease in initial abundances we have
calculated WM-Basic \citep{wmbasic} models of O stars and varied the
carbon abundance of the models, decreasing the mass fraction of carbon
by factors of 0.5, 0.1 and 0.01. We then measure the equivalent width
of the ultraviolet C\,{\sc IV} line as described by \citet{crow2}, and
compare it to that measured at a standard Solar-scaled carbon
abundance. The results are presented in Table \ref{civwidths}. It is
apparent that there is not a linear trend with metallicity, but rather
that the greatest effect is seen when $Z=0.004$ (i.e. significantly
sub-Solar, but not at the lowest metallicity). Here even a small
decrease in the carbon abundance decreases the C\,{\sc IV} EW by a
substantial fraction. At higher metallicities much more carbon must be
removed before any great effect is seen. At the lowest metallicities
there is already too little carbon in the atmosphere to produce a
strong C\,{\sc IV} line and so any further reduction has a minimal
effect.

\begin{figure}
\includegraphics[angle=0, width=84mm]{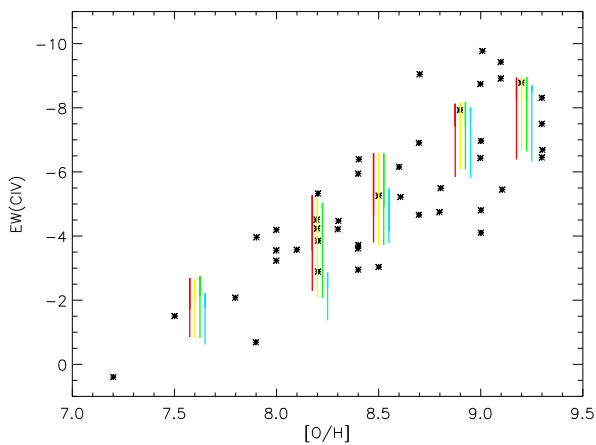}
\caption{The predicted equivalent width of C\,{\sc IV} at different
  metallicities measured by [O/H]. Lines represent BPASS
  models assuming constant star formation with the vertical extent due
  to age. The points are observations from \citet{crow2}. Red
  lines are for single stars, and yellow for binary models, the
  green lines indicate binary models including QHE and the blue lines are for
  binary models with the C\,{\sc IV} abundance decreased by a factor
  of four.}
\label{civew}
\end{figure}

To include the effect of varying composition in our spectral synthesis
we use the results in Table \ref{civwidths} to reduce the C\,{\sc IV}
EW in each model atmosphere according to its carbon abundance. We found
the accounting for the decrease of the carbon abundance predicted by
stellar models gave rise to only a tiny change in the C\,{\sc IV}
EW. Therefore we applied the maximum decrease in the carbon to oxygen
ratio found by \citet{erb} at $z\sim2$ so as to investigate the variation
of the initial carbon/oxygen abundance ratio. We present two sets of
models, one with normal Solar-scaled carbon abundances and a second
with the carbon abundance decreased by a factor of four. We only
perform this decrease for main-sequence OB stars when we use the OB
model atmospheres of \citet{crow}. The result is general reduction in
the strength of the C\,{\sc IV} absorption line as measured by the
equivalent width. As Figure \ref{civew} demonstrates this reduction
can go to explain some of the scatter in the observed EWs of C\,{\sc
  IV} in nearby galaxies, although the majority of the scatter
is due to variations in their star formation history.

\begin{table}
 \caption{The relative decrease in the C\,{\sc IV} line flux at
   1550\AA\ when the carbon abundance is decreased by a factors of 2,
   10 and 100. The values were estimated from \textsc{WM-BASIC}
   models \citep{wmbasic} with varying carbon abundances and constant
   overall metallicity. $X_{c}$ is the initial carbon mass fraction at
   each metallicity.}
  \begin{tabular}{@{}lccc}
    \hline
          &            EW(C\,{\sc IV}, $0.5 X_{c}$) &       EW(C\,{\sc IV}, $0.1 X_{c}$) &       EW(C\,{\sc IV}, $0.01 X_{c}$)\\
   $Z$    &             /EW(C\,{\sc IV}, $X_{c}$) &        /EW(C\,{\sc IV}, $X_{c}$) &        /EW(C\,{\sc IV}, $X_{c}$)  \\

\hline
0.001  & 0.828 & 0.692 &0.608\\
0.004  & 0.798 & 0.254 &0.100\\
0.008  & 0.931 & 0.592 &0.129\\
0.020  & 0.960 & 0.833 &0.295\\
\hline
\end{tabular}
 \label{civwidths}
\end{table}

\begin{figure}
\includegraphics[angle=0, width=84mm]{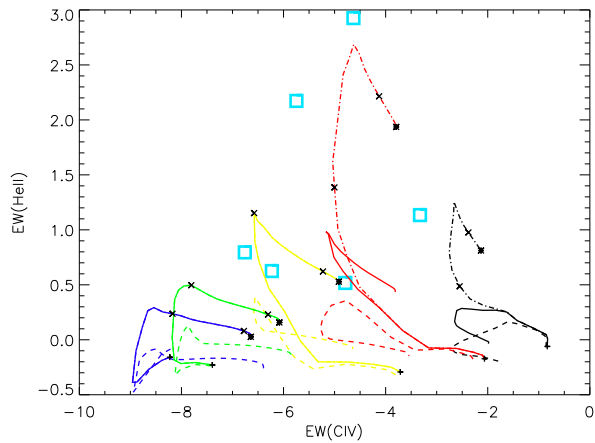}
\caption{How the strength of the He\,{\sc II} and C\,{\sc IV} line
  vary with age and metallicity for various BPASS stellar population
  models. The solid lines show binary models, the dash-dotted lines
  are for binary models including QHE and the dashed lines for single
  stars. Black indicates models with $Z=0.001$, the red lines have
  $Z=0.004$, the yellow line are for $Z=0.008$, the green line for
  $Z=0.020$ and the blue line for $Z=0.040$. On the binary model
  tracks, plus signs indicate the values at 1 Myr, with crosses along
  the line indicating ages of 10 Myr and 100 Myr , then the asterisk
  indicating the age of 1 Gyr. The light blue boxes indicate the EWs
  measured from the observed spectra described in Section
  \ref{sec:comp_spect_obs}.}
\label{civheii}
\end{figure}

\subsection{Producing a total synthetic population spectrum}

The procedure outlined above yields a synthetic spectrum appropriate
to each time-step of a stellar evolution model. We can then combine the
spectra for each star together to produce the integrated spectrum for
a synthetic stellar population. To do this we use the initial mass
function described by \citet{kroupa2002}. This uses an IMF power-law
slope of -1.3 between 0.1 and 0.5\,M$_{\odot}$, and a slope of -2.35
from 0.5 to 120\,M$_{\odot}$.

Finally in our spectral synthesis we include the contribution from
nebular emission. In star-forming galaxies, interstellar gas is
ionised by the stellar continuum emitted blueward of 912\AA, and upon
recombination it emits a nebular continuum. Neglecting this emission
would lead to an incorrect estimate of the equivalent widths of
emission lines and incorrect broad-band colours \citep{zack,molla}. We
use the radiative transfer program \textsc{Cloudy} \citep{cloudy} to
produce a detailed model of the output nebular emission spectrum
excited by our stellar spectra. The model output is sensitive to the
chosen geometry, inner radius and composition of the gas used in the
code. The details of our illustrative nebular emission model are
identical to those in \citet{ES09}, and we output the final continuum
and line strengths for use in our synthetic spectra.
%

\begin{figure*}
\includegraphics[angle=0, width=84mm]{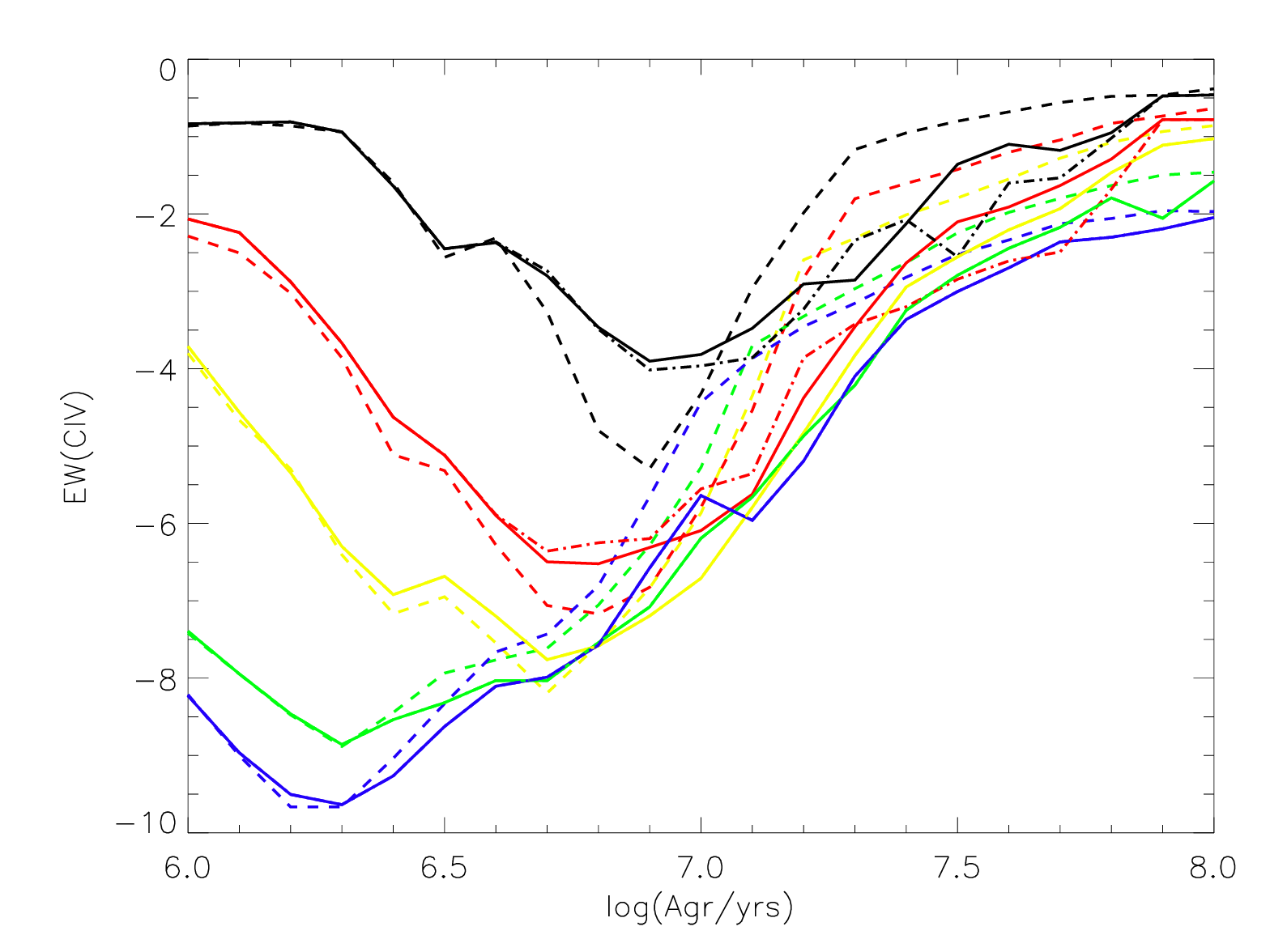}
\includegraphics[angle=0, width=84mm]{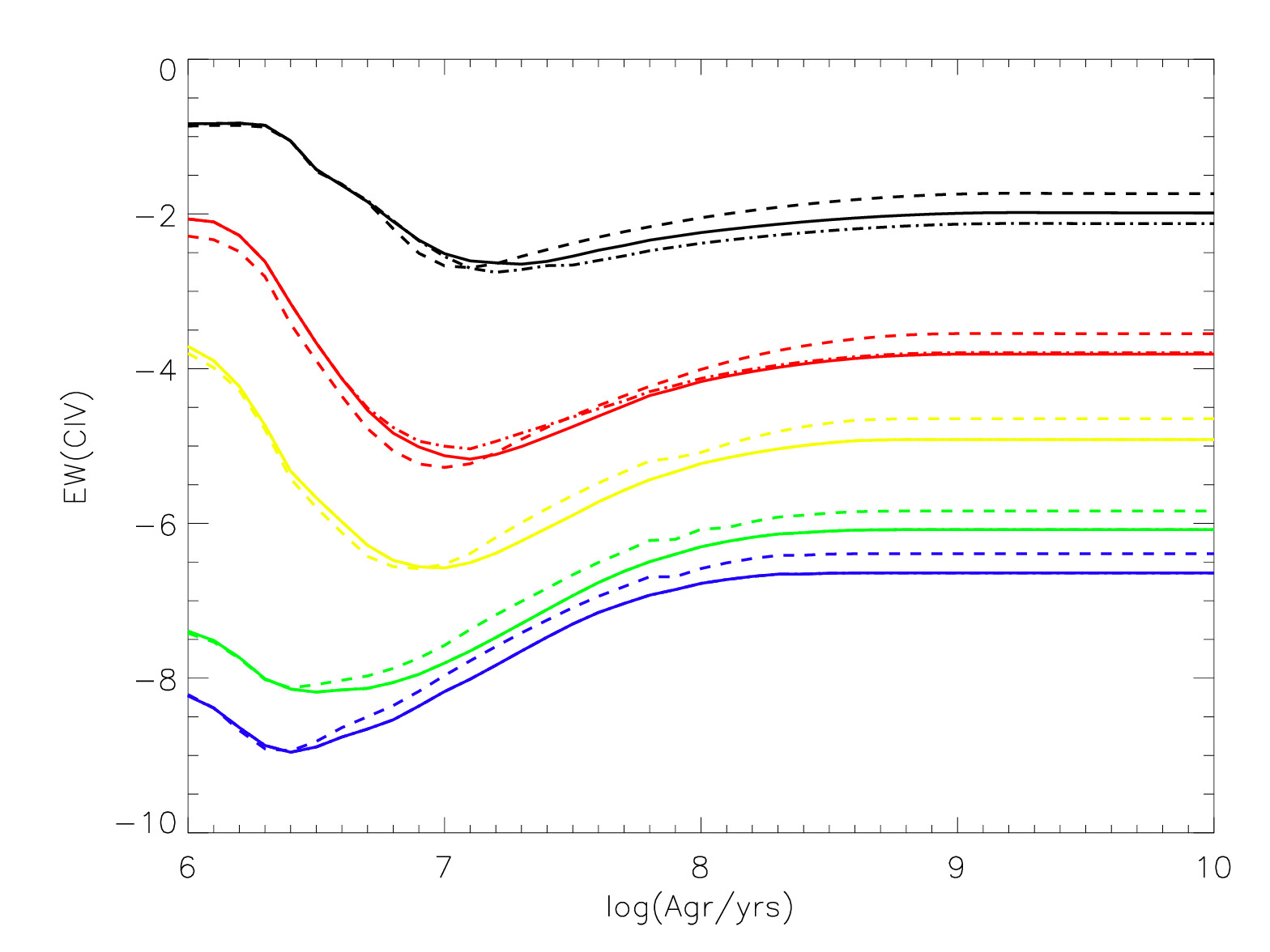}\\
\includegraphics[angle=0, width=84mm]{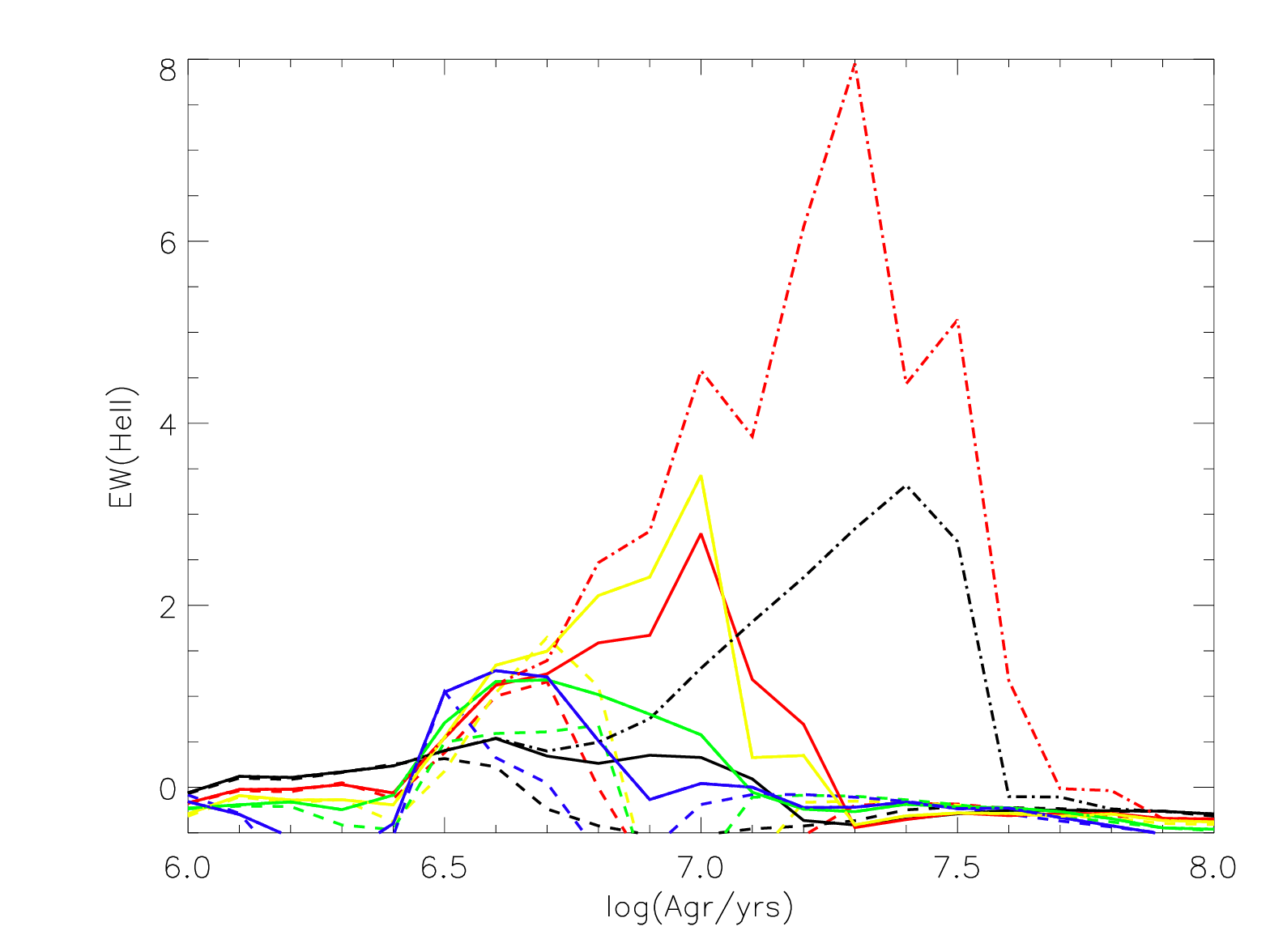}
\includegraphics[angle=0, width=84mm]{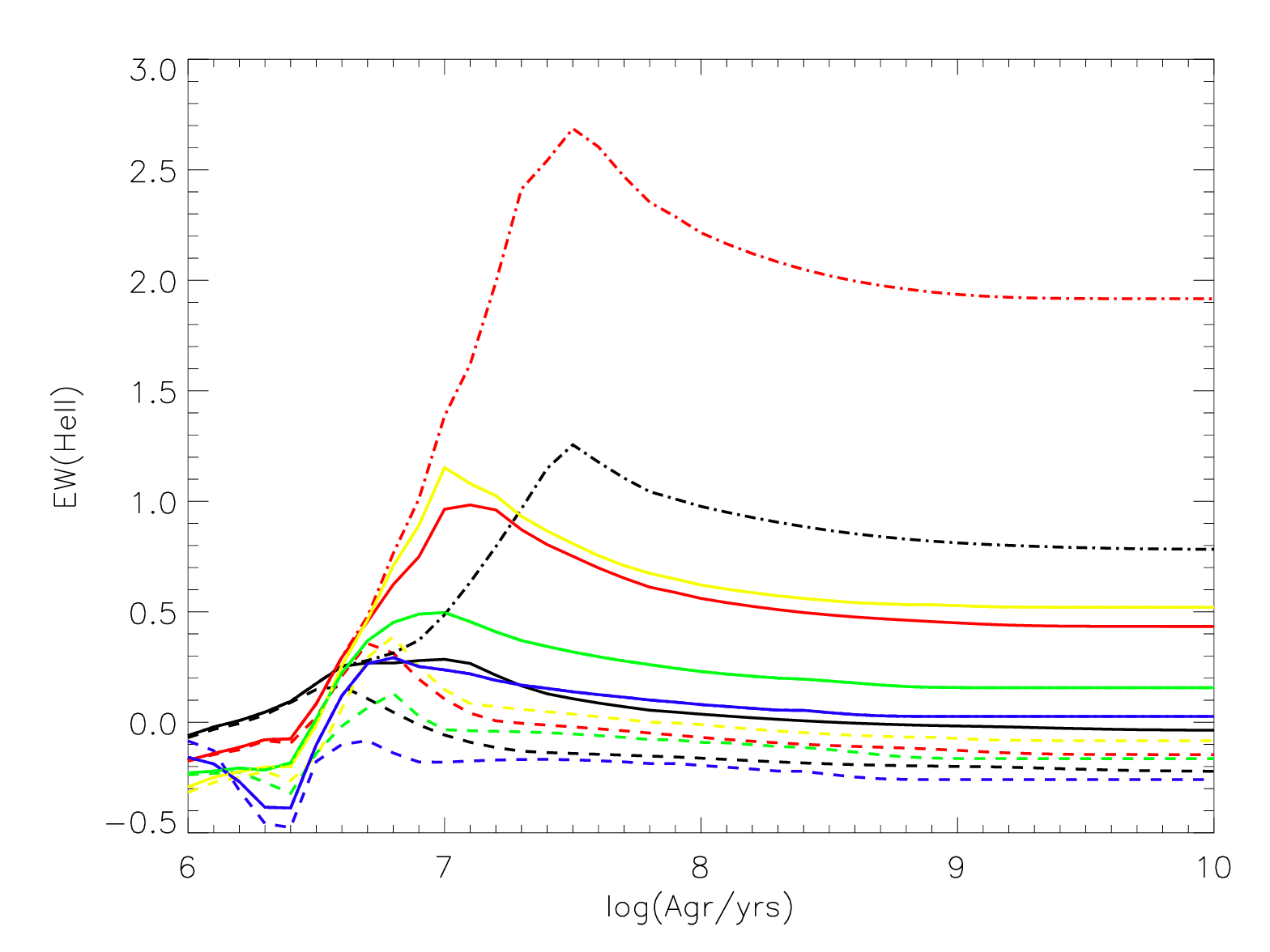}
\caption{The predicted equivalent widths for the C\,{\sc IV} and
  He\,{\sc II} stellar wind lines (upper and lower panels
  respectively). The panels in the left-hand column assume an
  instantaneous burst of star formation. The panels in the right
  column assume a constant star formation rate. The solid lines are
  for the standard binary models, the dashed line are those from
  single stars models and the dash-dotted lines for binary models
  including QHE. The black lines are for $Z=0.001$, the red lines for
  $Z=0.004$, the yellow line for $Z=0.008$, the green line for
  $Z=0.020$ and the blue line for $Z=0.040$.}
\label{modelew}
\end{figure*}

\section{Predicted Equivalent Widths}
\label{sec:pred-equiv-widths}

We consider in this work the C\,{\sc IV} and He\,{\sc II} spectral
lines, both of which exhibit a broad component whose strength is
determined by the stellar spectrum. We have already shown in Figure
\ref{civew} that model predictions for the C\,{\sc IV} line agree with
those deduced from observations. Conventional wisdom has been that the
magnitude of He\,{\sc II} line EW decreases in tandem with the equivalent width of
C\,{\sc IV} as shown by \citet{brinchmann1}. This is because it is
assumed that at lower metallicity there are fewer WR stars with
shorter lifetimes to contribute to the spectral features. However, as
shown in \citet{ES09}, the inclusion of binaries make this predicted
relation less clear-cut.

In Figure \ref{civheii} we show how the C\,{\sc IV} EW, which does
decrease with metallicity, relates to the He\,{\sc II} line strength
for our different model populations. When only single star populations
are considered, the strength of the He\,{\sc II} line is largely
independent of C\,{\sc IV} line strength, and, surprisingly, peaks at
around $Z=0.008$ rather than at the highest (Solar and super-Solar)
metallicity. This is because absorption lines in B star spectra
decrease the apparent strength of He\,{\sc II} at higher
metallicity. Stellar populations including binaries reach higher EWs
than the single star populations because of their greater number of WR
stars (produced due to mass-transfer events in the binary
systems). The highest EWs are only possible, however, when the effect
of QHE is included in a binary population. The production of long
lived hydrogen burning stars that are observed as WR stars boosts the line
strength at the lowest metallicities.

The same trends are evident in Figure \ref{modelew}. The highest
He\,{\sc II} EWs (i.e. strongest line emission) arise at later ages in
QHE models than for the single star populations. This is because the
additional WR stars produced from binary evolution and QHE are
typically from lower mass stars that would not normally become WR
stars and take of order 10$^7$ years to evolve to this state. Also in
Figure \ref{modelew} we show how both C\,{\sc IV} and He\,{\sc II}
lines vary assuming an instantaneous burst of star formation. This model
may be more appropriate at the highest redshifts ($z>5$) where the stellar
populations appear to be short-lived, dramatic events with little fuel
available for ongoing star formation \citep[see][]{2010MNRAS.408L..31D}.  

In this instantaneous star formation case, larger EWs are possible for
He\,{\sc II} than in the constant star formation model. This is only
because there are no younger stars contributing to the continuum and
reducing the relative significance of WR line flux in the equivalent
width (i.e. line-to-continuum flux ratio). However the He\,{\sc II}
emission lines typically observed in Lyman-break galaxies with good
spectroscopic data at $z\sim2-3$ do not require these extreme EWs, but
remain in agreement with those predicted by our constant star
formation models.

\section{Applications to the rest-frame UV spectra of Lyman-Break Galaxies}
\label{sec:comp_spect}

\subsection{Observations}
\label{sec:comp_spect_obs}

We have collated the spectra of five Lyman Break
galaxies at $z=2-3$ for which high quality rest-UV spectroscopy has been
published, four are gravitationally-lensed galaxies and the
other spectrum is the result of a deep, 12 hour integration. The strong magnification of
distant sources by the gravitational effects of intervening matter
allows the study of individual galaxies, rather than the necessary
blurring of any variation in a population through use of a
composite. Before comparing these spectra to models it is useful to
consider the spectra in comparison with each other and the earlier LBG
composite of \citet{shapley}.

\begin{figure}
\includegraphics[angle=0, width=84mm]{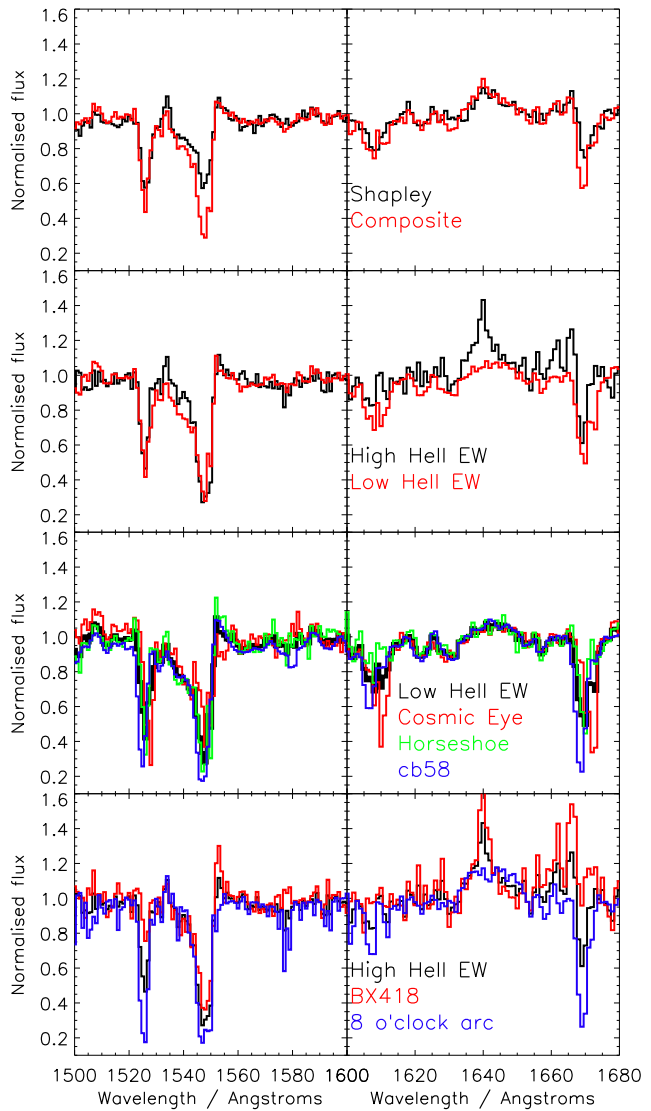}
\caption{Comparison of the different observed spectra. In the upper
  two panels we compare the Shapley composite with a composite of the
  lensed galaxies, below this we compare composites of the high and
  low He\,{\sc II} galaxies, in the final we panels we compare each
  lensed spectra to the relevant composite.}
\label{obsstacks}
\end{figure}

The data, in order of publication, comprise rest-frame ultraviolet
spectra for MS 1512-cB58 \citep[hereafter cB58,][]{cb58paper} at
$z=2.72$, the Cosmic Horseshoe \citep{2007ApJ...671L...9B,horse} at
$z=2.38$, the 8 o'clock arc \citep{8arc,2010A&A...510A..26D} at
$z=2.73$, the Cosmic Eye \citep{quid1} at $z=3.07$ and Q2343-BX418
\citep[hereafter BX418,][]{erb} at $z=2.30$. In Figure \ref{obsstacks}
we combine the normalised rest-frame ultraviolet spectra to form a
composite of the lensed galaxies and compare it to the $z\approx3$
Lyman-break galaxy composite of \citet{shapley}. As the upper panels
of the figure illustrate, there appears to be a remarkable qualitative
agreement between our small composite and that based on $\sim900$
individual galaxies at lower signal to noise. This suggests that,
taken as a group, these Lyman-break galaxy examples are
representative of the more numerous photometrically-selected
sample. The primary difference between the two composites is that the
C\,{\sc IV} absorption feature (in both its broad stellar wind and
narrow nebular components) is somewhat deeper in the lensed galaxies
than, on average, in their unlensed counterparts.

However, even this small composite does not tell the whole story. When
studying the C\,{\sc IV} and He\,{\sc II} profiles of the individual
spectra it is apparent that the lensed sources form two loose groups
that can be separated by the strength of their He\,{\sc II} emission
line. One set contains strong He\,{\sc II} stellar emission with
equivalent widths (EWs) of 2\AA\ or more \citep{8arc,erb}, while a
second group presents only weak or absent He\,{\sc II} emission
\citep{cb58paper,horse,quid1}. We compare the mean spectra of these high
and low He\,{\sc II} EW sets in the second set of panels in Figure
\ref{obsstacks}. Not only does the He\,{\sc II} strength vary
significantly between these two cases, but the C\,{\sc IV} profile is
also deeper in the weak-He\,{\sc II} case, indicative of a higher
carbon abundance.

In the final two sets of panels we show the individual spectra of the
lensed galaxies. It is clear that the spectra of sources in the low
He\,{\sc II} EW category are remarkably uniform. There is very little
difference between any of the observed spectra.  The two sources with
high He\,{\sc II} EW spectra are also similar to one another and
differ primarily in the strong (relatively narrow)
nebular emission contributing to the He\,{\sc II} line of the BX418
spectrum. \citet{erb} fit a dual Gaussian model to the He\,{\sc II}
emission line in this source and estimate that 20 per cent of the line
flux is contributed by nebular emission. Neglecting the nebular
contribution, the two spectra have identical He\,{\sc II} equivalent widths.

This qualitative comparison suggests that there may be evidence for two
distinct classes of $z\sim3$ LBGs, separated by the strength of the broad
He\,{\sc II} stellar emission line. Furthermore, it would appear
likely that both types of galaxies are represented within the
photometrically-selected galaxy sample contributing to the Shapley et
al. spectral stack.

\subsection{Modelling observed $z\sim2-3$ galaxies}
\label{sec:comp-lbgs}

\subsubsection{Shapley et al (2003) composite}

Given the new elements in our stellar population synthesis code, it is
informative to perform a direct comparison between the spectral
features predicted by the refined BPASS models and those observed in
the six Lyman-break galaxy spectra (five individual, one composite)
described in Section \ref{sec:comp_spect_obs}.  This comprehensive
study enables us to gain further insight into both the accuracy of our
models and the implied properties of the high-redshift galaxies
themselves.

\begin{figure}
\includegraphics[angle=0, width=84mm]{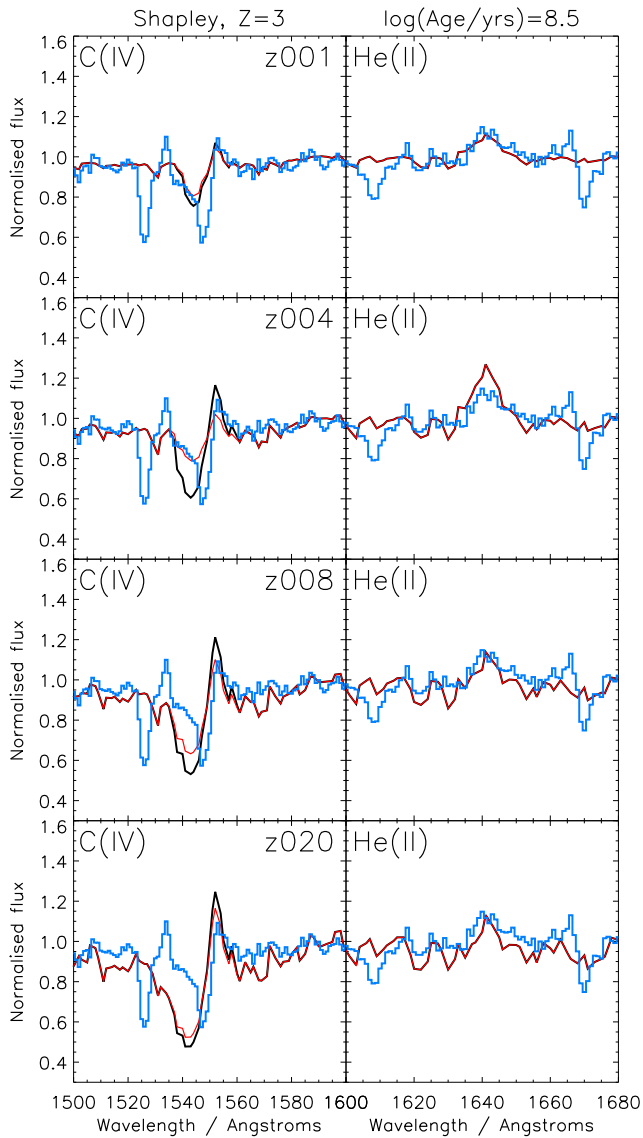}
\caption{A comparison between spectral features in the observed
  composite rest-frame UV spectra LBG of \citet{shapley} in blue and our binary
  population models including QHE, shown with and without a reduction
  in the carbon abundance in red and black respectively. Left-hand
  panels are for C\,{\sc IV} and panels on the right for the He\,{\sc
    II} line. The model spectra are for continuous star-formation
  lasting $10^{8.5}$ yrs. We show four different metallicities, from
  top to bottom $Z=0.001$, 0.004, 0.008 and 0.020.}
\label{fig:shapley}
\end{figure}

\begin{figure}
\includegraphics[angle=0, width=84mm]{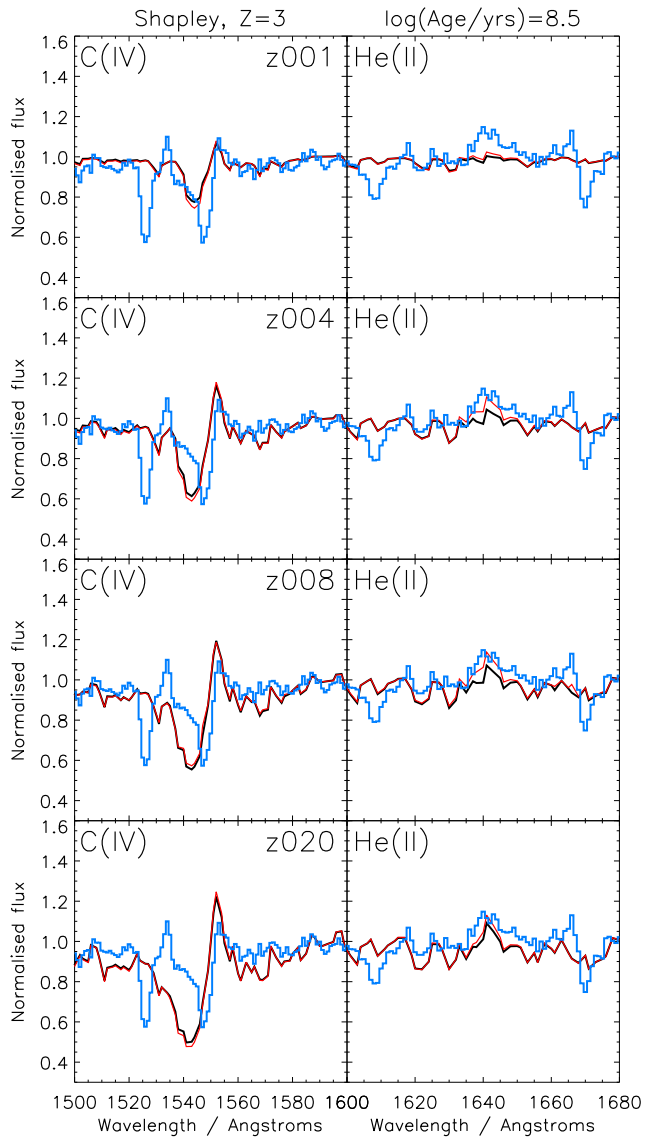}
\caption{Similar to Figure \ref{fig:shapley} but showing a pure single-star
  population in black and a binary population model excluding QHE in
  red.}
\label{fig:shapley2}
\end{figure}

In Figure \ref{fig:shapley} we present a comparison between the
composite spectrum of $\sim$900 LBGs with a mean redshift of three
\citep{shapley} and our model spectra, taken log(age/yrs)=8.5 into
an ongoing star formation episode \citep[found to be the median age
  and star formation history of this sample by][]{shapley01}. In the
original paper presenting this rest-UV composite, the authors found it
difficult to simultaneously reproduce the line strengths of the
C\,{\sc IV} and He\,{\sc II} line without resorting to models with
extreme IMFs. Later \citet{brinchmann1} was able to predict the
required EW for the He\,{\sc II} line but only at a lower metallicity
than suggested by other measurements of the sample.

We see in Figure \ref{fig:shapley} that the C\,{\sc IV} line in the
LBG composite is relative weak compared to our Solar metallicity models. Only
at sub-Solar metallicities ($Z$=0.001-0.004) can the models reproduce
the broad component of the observed P-Cygni wind profile shape. However at the higher of
these metallicities, the line profile can only be reproduced if the
C/O ratio is a quarter of that in the standard Solar abundances. This
is close to the depletion of this abundance ratio by a factor of 0.2
(relative to Solar) found by \citet{shapley} for the nebular gas in this
source. This implies that the stars we are observing likely formed
from the same gas that still surrounds them at the epoch of observation.

For completeness, we compare the composite spectrum to our theoretical
single star and non-QHE binary evolution models in Figure
\ref{fig:shapley2}. There is little difference in the C\,{\sc IV} line
relative to the models incorporating QHE shown in Figure
\ref{fig:shapley}. However both our standard single star and earlier
binary models have very weak He\,{\sc II} lines compared to those
spectra where we apply a prescription for QHE, and under-predict the
observed line strength. This indicates that accounting for binary
stellar evolution alone cannot entirely explain the observed He\,{\sc
  II} profile, but rotation in binaries may play an important role in
producing the observed spectra.

Because the spectrum analysed here is a composite of a number of
different galaxies we can only attempt to determine a mean or range of
metallicities. A qualitative analysis suggests an allowed metallicity
range of $0.001 < Z < 0.008$ for the `typical' Lyman-break galaxy at
$z\sim3$. This corresponds to a $ 7.6 < [O/H] < 8.5$ (or $ 0.1 <
Z/Z_{\odot} < 0.6$ where $[O/H]_{\odot}=8.7$), and is consistent with
the previous measurements determined by \citet{shapley} and
\citet{2004ApJ...615...98R}.

\subsubsection{Individual galaxies}

\begin{figure}
\includegraphics[angle=0, width=84mm]{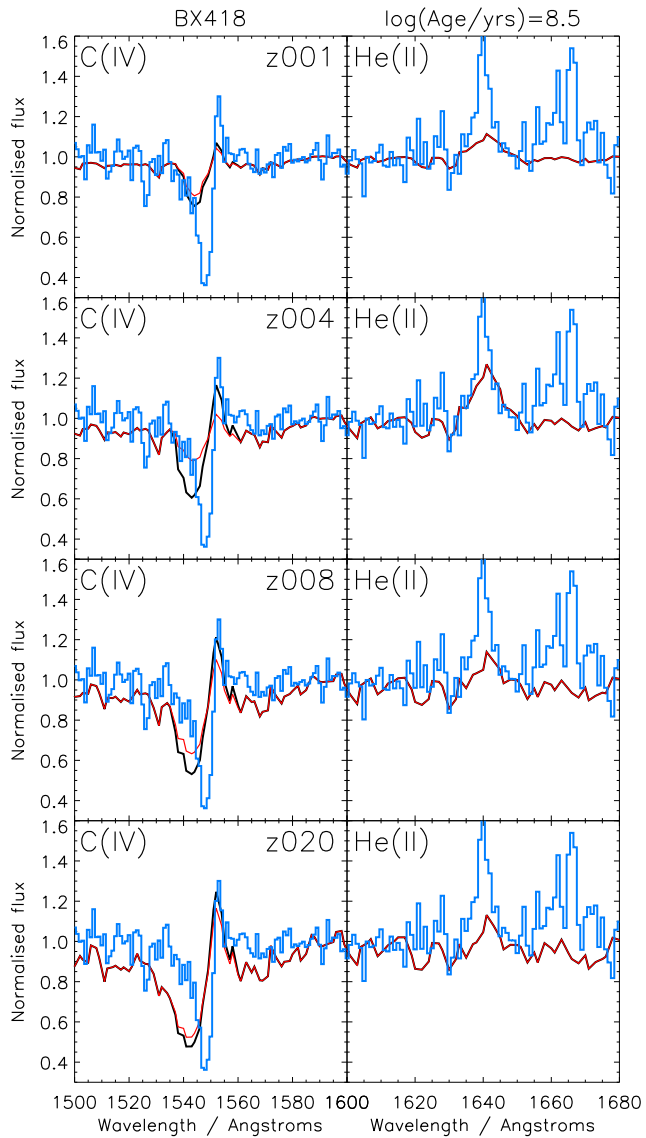}
\caption{Similar to Figure \ref{fig:shapley} but for the observed UV
  spectrum of BX418 \citep{erb}.}
\label{bx418}
\end{figure}

\begin{figure}
\includegraphics[angle=0, width=84mm]{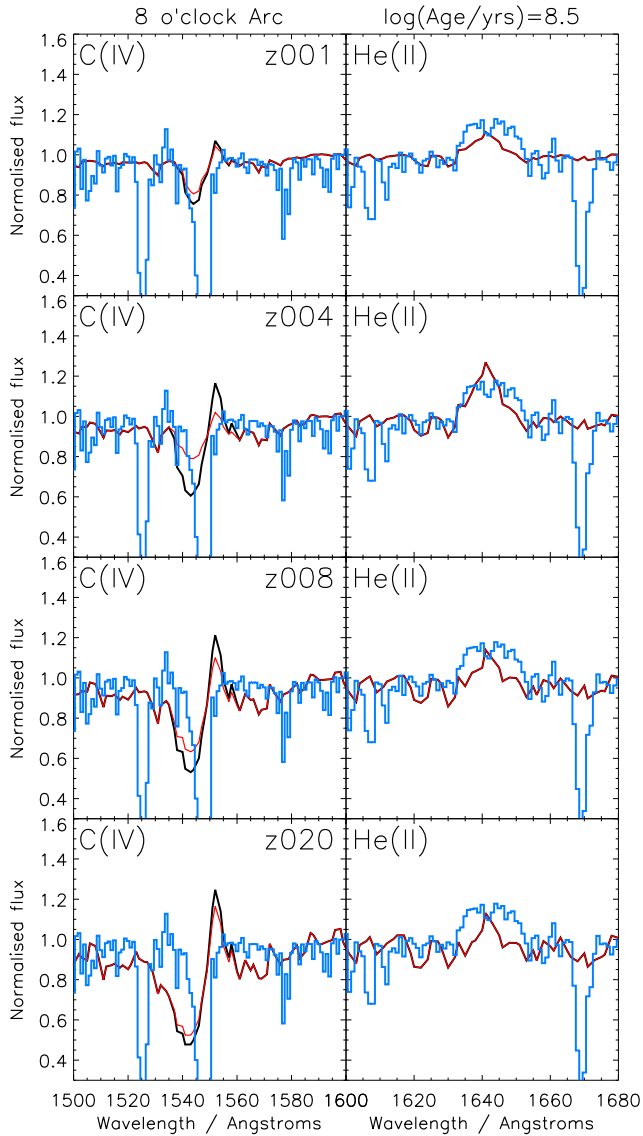}
\caption{Similar to Figure \ref{fig:shapley} but showing the observed UV
  spectrum of the 8 o'clock arc \citep{8arc}.}
\label{8oclockarc}
\end{figure}

\begin{figure}
\includegraphics[angle=0, width=84mm]{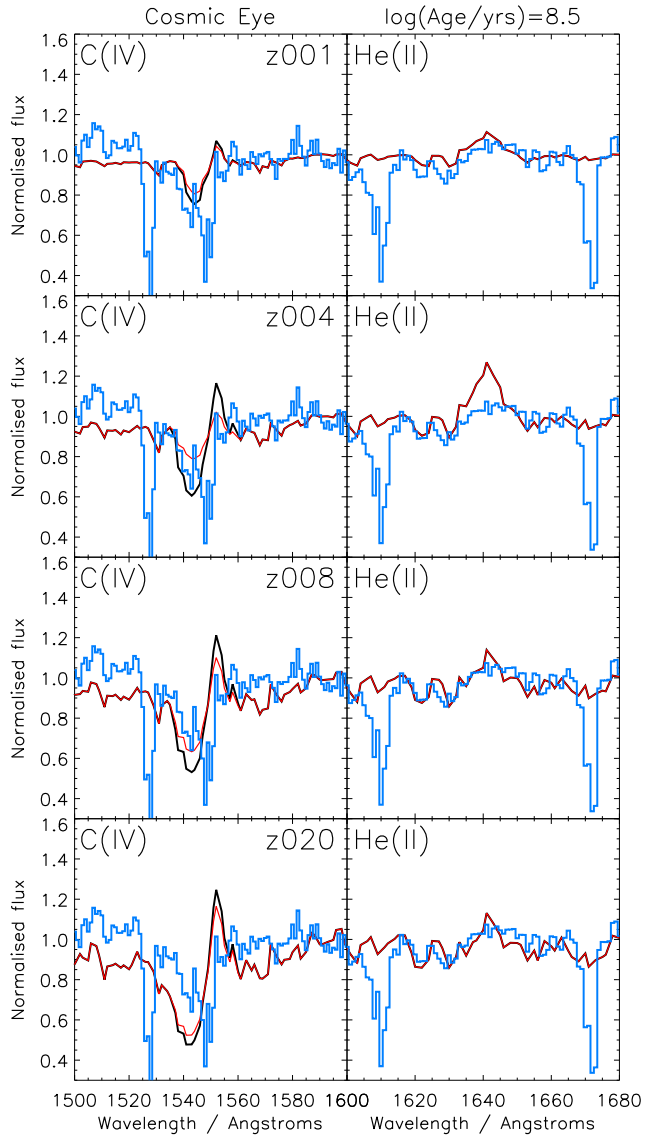}
\caption{Similar to Figure \ref{fig:shapley} but for the observed UV
  spectrum of the Cosmic Eye \citep{quid1}.}
\label{eye}
\end{figure}

\begin{figure}
\includegraphics[angle=0, width=84mm]{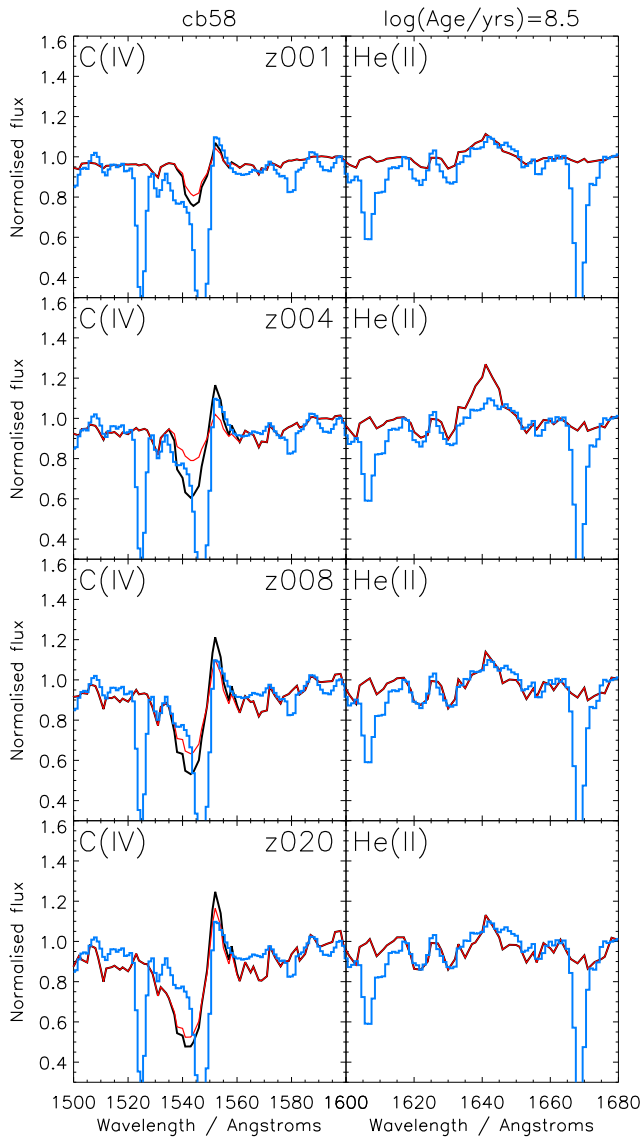}
\caption{Similar to Figure \ref{fig:shapley} but for the observed UV
  spectrum of cB58 \citep{cb58paper}.}
\label{cb58}
\end{figure}

\begin{figure}
\includegraphics[angle=0, width=84mm]{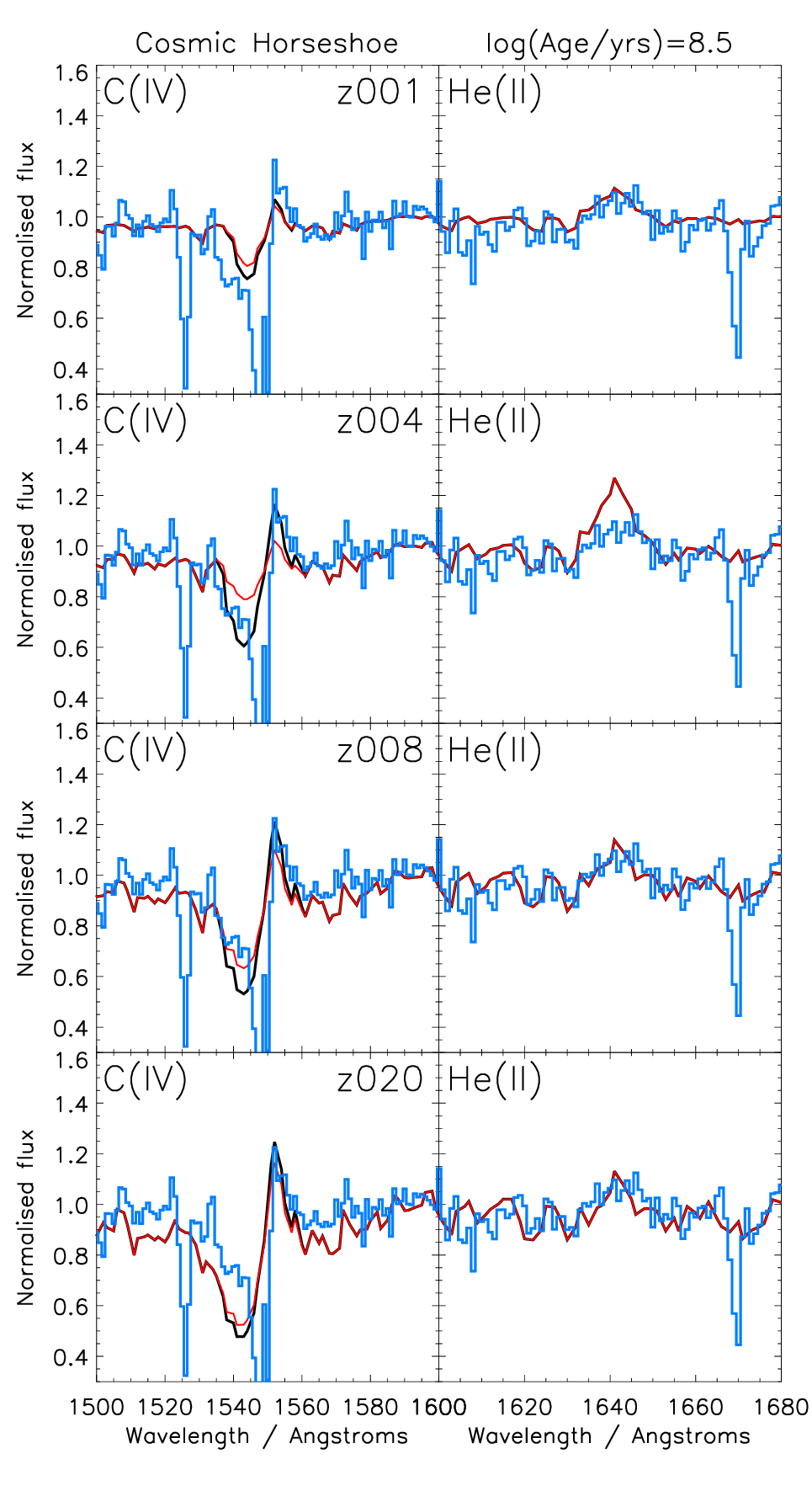}
\caption{Similar to Figure \ref{fig:shapley} but for the observed UV
  spectrum of the Cosmic Horseshoe \citep{horse}.}
\label{horse}
\end{figure}

We address the spectra of the individual lensed galaxies compiled here
starting with the sources with the lowest inferred metallicity, based
on our model comparisons. In each case, we make the comparison at a
constant log(age/years)=8.5 (age=320\,Myr), based on the median
star-formation derived age of the $z\sim3$ Lyman-break galaxy
population as a whole. We note that, as figure \ref{modelew}
indicates, there is little evolution in the model strengths of our lines
of interest at stellar population ages above $\sim$100\,Myr.

The UV spectrum of the $z=2.3$ galaxy BX418 from \citet{erb} (shown in
Figure \ref{bx418}) has a very weak C\,{\sc IV} stellar wind line with
a P Cygni profile and strong and broad He\,{\sc II} emission. There is
clearly a narrow nebular component superposed on the broader He\,{\sc
  II} stellar emission line. This indicates a very hard ionising
spectrum present within the galaxy. The He\,{\sc II} line implies
that the metallicity must be low, especially in carbon, suggesting a
metallicity range of $0.001$ to $0.004$ (i.e. significantly
sub-Solar). However the more prominent C\,{\sc IV} profile implies a
slightly higher (but still sub-Solar) carbon abundance in this
source. As is the case for the Shapley et al composite, matching 
both lines simultaneously requires the presence
of both a reduced relative carbon abundance and QHE in the population
synthesis model.

In Figure \ref{8oclockarc} we consider the 8 o'clock arc, a $z=2.7$
lensed Lyman break galaxy, with a rest-frame UV spectrum obtained by
\citet{2010A&A...510A..26D}. This galaxy has reported metallicities
higher than others in this sample (Solar, or near-Solar) and has also
been reported as relatively massive and dusty
\citep[M$_\ast\sim4\times10^{11}$\,M$_\odot$, E(B-V)$\sim$0.67, see
][]{finkel,2010A&A...510A..26D}. In this context, its rest-UV spectrum
is somewhat surprising.  As in the case of BX418, the 8 o'clock arc
has a very weak C\,{\sc IV} stellar wind line (the absorption is
dominated by the narrow nebular component), but strong and broad
He\,{\sc II} emission, without any obvious nebular component. The lack
of a strong C\,{\sc IV} wind line indicates that the preferred
metallicity in the regions dominating the rest-UV emission
must be low, especially in carbon, suggesting a
metallicity range of $0.001$ to $0.004$, with higher metallicities
within this range requiring a reduced [C/O] ratio. The strong He\,{\sc
  II} bump suggests metallicities at the high end of the allowed range
to reproduce this line. At the lower metallicities fewer WR stars are
produced, and weaker line emission in the WR stars also reduces the
strength of the line observed. This He\,{\sc II} line strength is also
only reproducible with QHE in the stellar population.

Based solely on the stellar-dominated spectral features, the
similarity between the rest-UV spectrum of this galaxy to that of
BX418 (shown in Figure \ref{obsstacks}) would suggest they are of a
similar metallicity, with the deeper absorption features hinting at a
slightly more metal rich environment in the 8 o'clock arc. This is
somewhat in contradiction to the higher derived metallicity range
found by \citet{finkel} and \citet{2010A&A...510A..26D} based on
nebular ionization models and older stellar population models. It is
possible that any discrepancy may arise in part from the different
regions probed by these different measurements. If so, it would imply
that the youngest star formation (and hence the massive stars that
dominate the He\,{\sc II} and C\,{\sc IV} lines under consideration
here) is taking place in regions of lower metallicity than the more
metal-rich gas elsewhere in the galaxy. It is, of course, dangerous to
over-interpret weak features in data with this signal to
noise. However, the flattened line profile of the He\,{\sc II}
emission may support the hypothesis that metal-rich gas is present to
some degree in this system, if interpreted as the superposition of a
narrower (perhaps nebular) absorption line on the peak of the broad
stellar emission feature.
 
Interestingly, \citet{cresci} have identified three star-forming
galaxies at a similar redshift that show evidence for an inverted
metallicity gradient (i.e. central star forming regions with lower
metallicity than outlying regions) which may arise from the central
accretion of cold gas. Like the 8 o'clock arc, all three are
relatively massive (M$_\ast\sim10^{11}$\,M$_\odot$) and they also show
dynamics that suggest rotational support rather than merger-driven
star formation (which may be more dominant in lower-mass UV-luminous
systems). A cold-accreting system of this kind is likely to display
the same discrepancy between metallicities determined from the
youngest stars and from other methods. Cold-mode accretion has been
hypothesised as a dominant galaxy formation mechanism at the highest
redshifts \citep{2009Natur.457..451D}, although this hypothesis
remains controversial \citep[see e.g.][]{2010ApJ...717..289S}. While
such a case has not been proven in the 8 o'clock arc, it may prove an
interesting target for detailed analysis with integral field
spectroscopy, to determine whether there is any evidence for this
scenario.

The remaining three galaxies differ to those already discussed in that
they do not require the simultaneous invocation of low
metallicity, reduced carbon abundance and QHE in their synthetic
stellar population to explain the spectral features under
consideration, but rather display some or none of these features.

The $z=3.1$ Cosmic Eye spectrum \citep{quid1} is illustrated in Figure
\ref{eye} and is notably devoid of strong He\,{\sc II} emission. This
can be explained if the galaxy lies above the metallicity limit at
which QHE no longer occurs. The absorption section of the C\,{\sc IV}
P Cygni profile also implies a relatively strong metallicity somewhere
between $Z=0.004$ and $0.008$, with a reduced carbon abundance
required at the higher metallicity mass fraction. Higher metallicities
are ruled out by the relatively shallow absorption.

The remaining two examples, cB58 \citep[$z=2.7$,][]{cb58paper} and the
Cosmic Horseshoe \citep[$z=2.4$,][]{horse}, in Figures \ref{cb58} and
\ref{horse} respectively, tell a similar story. The lack of a
strong He\,{\sc II} line, yet apparent C\,{\sc IV} P-Cygni profiles,
suggest higher metallicities for these galaxies, above the range of
metallicities for galaxies with stronger He\,{\sc II}.

These 
results suggest that the existence of two populations of LBG
galaxies, defined by the presence or absence of broad He\,{\sc II}
emission as discussed in Section \ref{sec:comp_spect_obs}, may be
explained by intrinsic galaxy properties differentiated primarily by
the overall metallicity mass fraction of the youngest stellar
population (which dominates the UV emission and hence the two stellar
features under consideration here).  If
the metallicity mass fraction is low then QHE is possible in binaries
and produces more long lived Of and WR stars, which contribute to
the broad He\,{\sc II} line. By contrast, when the metallicity is high
QHE is no longer possible because stellar wind mass-loss breaks the
rotation of stars rapidly to prevent this unusual evolution, and
He\,{\sc II} emission is comparatively weak.
Importantly, in both sets of galaxies we see evidence for a sub-Solar
carbon to oxygen ratio in the weaker than expected C\,{\sc IV} P-Cygni profiles.

\begin{table*}
\caption{The stellar population model found to best fit each of the
  Lyman-break galaxies discussed in section~\ref{sec:comp-lbgs},
  indicating the preferred range of metallicity and whether either a
  reduced carbon to oxygen ratio or quasi-homogeneous stellar evolution
  models are required to explain the observed spectrum. We list the
  details for composite LBG spectrum (which demonstrates the broadest
  allowed metallicity range) and the observed samples with strong and
  weak He\,{\sc II} line emission as discussed in Section
  \ref{sec:comp-lbgs}.}
\label{lbgsinfo}
\begin{tabular}{lccccccc}

\hline 
         & \multicolumn{4}{|c|}{This Paper} & \multicolumn{2}{|c|}{Previous Work} \\
\hline
                     &           &                &   Reduced        & Require          &          &       &        \\
Spectrum             &  $12+\log(O/H)$    &  $Z$  &    [C/O]?         &  QHE?           &    $12+\log(O/H)$   & [C/O]  &  Notes    \\
\hline
$z=3$ composite      & 7.6--8.5  & 0.001--0.008       & Yes              & Yes               &  7.7--8.7   & $0.2\pm0.1$  &                         \\
\\
BX418                & 7.6--8.2  & 0.001--0.004       & Yes              & Yes               &  7.7--8.1   & 0.2 &  Nebular He\,{\sc II} contribution \\
8 o'clock arc        & 7.6--8.2  & 0.001--0.004       & Yes              & Yes               &  8.3--8.6   & --                  & \\
\\
Cosmic Eye           & 8.2--8.5  & 0.004--0.008       & Yes              & No                &  $\sim 8.3$   & --&  Higher $Z$ preferred \\
cB58                 & 8.2--8.5  & 0.004--0.008       & Yes              & No                &  $\sim 8.4$   & --&  Higher $Z$ preferred\\
Cosmic Horseshoe     & 8.2--8.5  & 0.004--0.008       & Yes              & No                &  $\sim 8.4$   & --&  Higher $Z$ preferred \\
\hline
\end{tabular}
\end{table*}


\section{Conclusions}
\label{sec:conclusions}

In this paper we provide evidence that our new synthetic
stellar population models can explain the observed stellar emission
lines in the UV spectra of LBGs, with best fit models as summarised in
Table \ref{lbgsinfo}. We see evidence for two main types of LBG UV
spectra, differentiated by the presence or absence of broad He\,{\sc
  II} line emission, and explained as the result of varying
metallicity in the youngest stellar population. This leads to the
absence/presence of an extreme rotational effect (quasi-homogeneous
evolution) in the stellar binaries incorporated in our synthesis
models. We find three details that must be included to model the
spectra accurately: massive binary stars, QHE from the spin-up of the
secondary stars in binaries in the event of mass-transfer events, and
a reduced carbon-to-oxygen abundance ratio.

Our models support the inference that the sub-Solar [C/O] ratio
derived from nebula abundances in the $z=2.3$ galaxy studied in detail
by \citet{erb} may be wide-spread in this population, and that its
inclusion in synthesis models is important when fitting individual
spectral features. Depletion by a factor of approximately four is
consistent with the entirety of the sample discussed here.

In addition, we confirm our earlier work which suggests that binary
evolution has significant effects on the spectra of unresolved stellar
populations. The main effect of this is increased mass loss in binary
stars, and therefore an increase in the overall number of WR
stars. Furthermore the effects of rotation (QHE) within a binary
system can also lead to dramatic and hither-to unexpected results in
the derived model stellar spectra.

Given the near-ubiquity of sub-Solar carbon-to-oxygen ratios in this
sample, and the increasing import of QHE effects at low metallicity,
it is interesting to consider the possible implications for analysis
of Lyman-break galaxies at still higher redshift. This analysis has
focused on galaxies at $z\la3$. However similar sources, selected for
high rest-UV flux (and hence recent star formation) have now been
identified to $z=8$ and beyond. The detailed spectroscopic analysis of
these sources is still more difficult than that at $z=3$ due to a
combination of increased sky background at long wavelengths, decreased
detector sensitivity and increasing luminosity distance, and is not
considered further here. However, forthcoming facilities such as the
{\em James Clark Maxwell Telescope} (JCMT) and the KMOS spectrograph may make
analysis of the He\,{\sc II} line and other, longer wavelength
spectral signatures relatively straightforward at $z=4$ and above. At
the low, but non-zero, metallicities inferred at $z=4-8$
\citep[e.g.][]{2004ApJ...610..635A,2010MNRAS.409.1155D}, the effects
of depleted carbon abundance and QHE may have a significant effect on
observed spectral features and improved modelling in these areas will
be essential for accurate interpretation of spectra obtained.

As the case of the 8 o'clock arc makes clear, it is possible that the
youngest star formation is atypical of the galaxy as a whole, which
has implications for our understanding of the processes of galaxy
formation. If cold-mode accretion is indeed increasingly common at
high redshifts as has been suggested
\citep[e.g.][]{cresci,2009Natur.457..451D} then stellar emission and
absorption features may be an accessible indicator of this.  While the
limited sample discussed here is, of course, difficult to interpret,
the use of KMOS and the JCMT to build a sample of high redshift
sources with high quality rest-UV spectra may allow the identification
of galaxies for which the youngest stellar populations are anomalously
low in metallicity. Such sources may well prove interesting targets to
study as possible examples of the still hotly-debated cold-mode
accretion scenario.

\section*{Acknowledgements}

The authors would like to thank those who kindly provided their
spectra and made this paper possible, Miroslava Dessauge, Dawn Erb,
Anna Quider and Max Pettini. The authors would also like to thank the
anonymous referee for bringing the peculiar nature of the 8 o'clock
arc to their attention. The authors would like to thank Malcolm
Bremer, Max Pettini, Paul Crowther, Stephen Smartt, Nate Bastian and
Norbert Langer for useful discussions.  ERS acknowledges postdoctoral
research support from the UK Science and Technology Facilities Council
(STFC) for part of this work. JJE acknowledges support from the UK Science and Technology
Facilities Council (STFC) under the rolling theory grant for the
Institute of Astronomy.

\label{lastpage}
\bsp

\end{document}